\documentclass{WileyMSP-template}
\usepackage{amsmath,bm}
\usepackage{braket}
\usepackage{multirow}

\renewcommand{\vec}[1]{\mathbf{#1}}

\begin{document}

%\pagestyle{fancy}
%\rhead{\includegraphics[width=2.5cm]{vch-logo.png}}
%\rhead{\includegraphics[width=2.5cm]{}}

\title{Topological Approach of Characterizing Optical Skyrmions and 
Skyrmion Lattices}

\maketitle

% Author: Please give full first and last names for authors and include * after the name of all corresponding authors

\author{Amy McWilliam*}
\author{Claire M. Cisowski}
\author{Zhujun Ye}
\author{Fiona C. Speirits}
\author{J\"{o}rg B. G{\"o}tte}
\author{Stephen M. Barnett}
\author{Sonja Franke-Arnold}

% Affiliations: Please provide adacemic titles (Prof. or Dr.) for all authors where applicable, and include an institutional email address for all corresponding authors
\begin{affiliations}
A. McWilliam, C. M. Cisowski Z. Ye, F. C. Speirits, J. B. G{\"o}tte, S. M. Barnett, S. Franke-Arnold\\
School of Physics and Astronomy, University of Glasgow, Glasgow, UK\\
Email Address: a.mcwilliam.1@research.gla.ac.uk

\end{affiliations}

% Keywords: Please provide a minimum of three and a maximum of seven keywords, separated by commas


% Abstract should be written in the present tense and impersonal style (i.e., avoid we), and be at most 200 words long
\begin{abstract}

The Skyrmion number of paraxial optical Skyrmions can be defined solely via their polarization singularities and associated winding numbers, using a mathematical derivation that exploits Stokes's theorem. 
It is demonstrated that this definition provides a robust way to extract the Skyrmion number from experimental data, as illustrated for a variety of optical (N\'eel-type) Skyrmions and bimerons, and their corresponding lattices. 
This method generates not only an increase in accuracy, but also provides an intuitive geometrical approach to understanding the topology of such quasi-particles of light, and their robustness against smooth transformations.

\end{abstract}

% Text: Please use section headings and subheadings as specified below. For communications, all section headings apart from Experimental Section should be removed
% Please make the first reference to a display item bold: \textbf{Figure 1}
% Do not abbreviate Figure, Equation, etc.; display items are always singular, i.e., Figure 1 and 2.
% Equations are always singular, i.e., Equation 1 and 2, and should be inserted using the {equation} environment, not as graphics
% Please do not use footnotes in the text, additional information can be added to the Reference list.

%%%%%%%%%%%%%%%%%%%%%%%%%%
\section{Introduction}
%%%%%%%%%%%%%%%%%%%%%%%%%%

The concept of Skyrmions was proposed by Skyrme over 60 years ago \cite{Skyrme1961}, originally postulated to describe the topological structure of nucleons.
Since then these quasi-particles have been predicted and observed in a wide range of contexts, including string theory \cite{Sutcliffe2010},  Bose-condensates and atoms \cite{Khawaja2001,Parmee2022}, spintronics \cite{Rho2015}, magnetic media \cite{Bogdanov2020, Tokura2021,Back2020}, and more recently in plasmonics  and optics \cite{Donati2016,Tsesses2018, Du2019, Gao2020}. While the more familiar magnetic Skyrmions carry magnetic spin textures, optical Skyrmions are embedded in the polarization texture of complex vector light fields. Of particular interest to the experimentalist are 2D Skyrmions, sometimes called `baby Skyrmions', which can be  realised in paraxial beams, offering an easily accessible and re-configurable platform for the investigation of topological features and their propagation dynamics \cite{Sugic2021, Shen2022, Zhu2021,Lei2021,Pang2022}. The generation of topological states of light opens up new avenues for the controlled interaction of photons with material quasi-particles such as plasmons, phonons and excitons \cite{Rivera2020}. 

Experimentally, polarization textures and hence Skyrmions can be assessed by measuring the spatially varying reduced Stokes vector $\vec{S}(x,y)$ across the light profile, mapping the local polarization states onto the Poincar\'e sphere, 
just like the local spin of magnetic Skyrmions is mapped onto the Bloch sphere  \cite{Bogdanov1994}. 
The polarization texture itself can take on almost unlimited shapes, including N\'eel-type (hedgehog) and Bloch-type Skyrmions~\cite{Gao2020,Cuevas_2021,Shen2022,cisowski2022,Shen2022_b}, but the underlying topology is characterized by a single invariant, the Skyrmion number, $n$, which counts how many times $\vec{S}$ wraps around
the Poincar\'e sphere \cite{footnote1}.
%Note that while every Skyrmion beam is a Poincar\'e beam, the reverse does not hold: The mapping must obey specific mapping rules. This is intrinsic to the definition of the Skyrmion number, Equation~(\ref{Eq:SkyrmNum}) but becomes more evident from our geometric description in this letter. 
While the polarization structure of a beam may change upon propagation in free space and through unitary transformations, 
$n$ remains conserved. 

Both isolated optical Skyrmions and and more complex geometries with multiple optical Skyrmions can be generated
with state-of-the-art light-shaping technology. Their experimental identification in terms of their Skyrmion number, however, remains challenging. Previous work has relied either on a qualitative comparison between measured and ideal polarization profiles 
of the target Skyrmion, or evaluated the Skyrmion number 
from its analytical expression, which for a beam propagating along the $z$ direction, 
is defined as \cite{Nagaosa2013} 
\begin{equation}
    n =\frac{1}{4\pi}\int_A \vec{S}\cdot\bigg(\frac{\partial\vec{S}}{\partial x}\times  \frac{\partial\vec{S}}{\partial y}\bigg)dxdy,
    \label{Eq:SkyrmNum}
\end{equation}
where $\vec{S}=[S_1,S_2,S_3]^T$ is the spatially resolved normalized reduced Stokes vector, and $A$ is the entire $(x,y)$ plane.
The Skyrmion number is thus, by definition, a global property of a light beam. Evaluating Equation~(\ref{Eq:SkyrmNum}) poses two difficulties: Firstly, the experimentally accessible part of $A$ is limited by the numerical aperture of the system. Secondly, gradients are notoriously sensitive to noise, especially in low intensity regions, such as at large radial distances or in the vicinity of singularities, where the spatial derivative of fluctuating noise levels may overwhelm the signal. 

In this letter we derive and demonstrate an alternative, topological method to calculate Skyrmion numbers, which avoids products of polarization gradients and significantly increases precision (for low $n$) and accuracy (in the presence of noise). We evaluate $n$ for a variety of experimentally generated optical Skyrmions and Skyrmion lattices. Our method provides geometric insight that is missing from the surface integral representation: allowing us, for example, to interpret Skyrmion lattices as combination of individual Skyrmion structures at the lattice sites, 
rather than just providing an overall Skyrmion number.

%%%%%%%%%%%%%%%%%%%%%%%%%%
\section{Topological definition of the Skyrmion number}
%%%%%%%%%%%%%%%%%%%%%%%%%%

We start by deriving our topological definition from the integral definition of Equation~(\ref{Eq:SkyrmNum}). For paraxial beams, the Skyrmion number can be interpreted as the integrated flux of a Skyrmion field $\bm{\Sigma}$, sometimes known as the topological current, across the transverse plane, 
\begin{equation}
    n = \frac{1}{4\pi}\int_A \bm{\Sigma} \cdot d\vec{A},  \label{Eq.vIntegral1}
\end{equation}
where $\Sigma_i = \frac{1}{2}\varepsilon_{ijk}\varepsilon_{xyz}S_z (\partial_j S_x) (\partial_k S_y)$ and $\partial_i$ denotes differentiation with respect to $x_i$ \cite{Gao2020}. 
Here, $\vec{S}_R=(S_x, S_y, S_z)^T$ is a generalized Stokes vector that relate to the conventional Stokes vector through an arbitrary rotation, described by a 3D rotation matrix $R$, so that $ \vec{S}_R=\vec{S} $ \cite{Shen2022_b}.   

The Skyrmion field $\bm{\Sigma}$ is transverse ($\nabla\cdot\bm{\Sigma}=0$), hence it can be expressed as the curl of a vector field $\vec{v}$. 
Applying Stokes's theorem yields
\begin{equation}
    n = \frac{1}{4\pi}\int_A \nabla\times \vec{v} \cdot d\vec{A}  = \frac{1}{4\pi}\oint_C \vec{v}\cdot d\bm{l},
    \label{Eq.vIntegral}
\end{equation}
where $C$ is a suitable integration path across $A$, which excludes any singularities of $\vec{v}$. 
While $\vec{v}$ is not uniquely defined, a suitable expression in terms of the experimentally accessible Stokes parameters is 
\begin{equation}
    \vec{v}=-S_z \nabla \Phi, \quad \mathrm{where} \quad  \Phi =\arctan(S_y/S_x).
    \label{Eq.v}
\end{equation}
Equation~(\ref{Eq.v}) can be derived by recalling the Mermin and Ho relation \cite{Mermin1976}, but for our purposes it is sufficient that its curl is indeed the Skyrmion field $\bm{\Sigma}$. 
\begin{figure}[htbp]
    \centering
    \includegraphics[width=7cm]{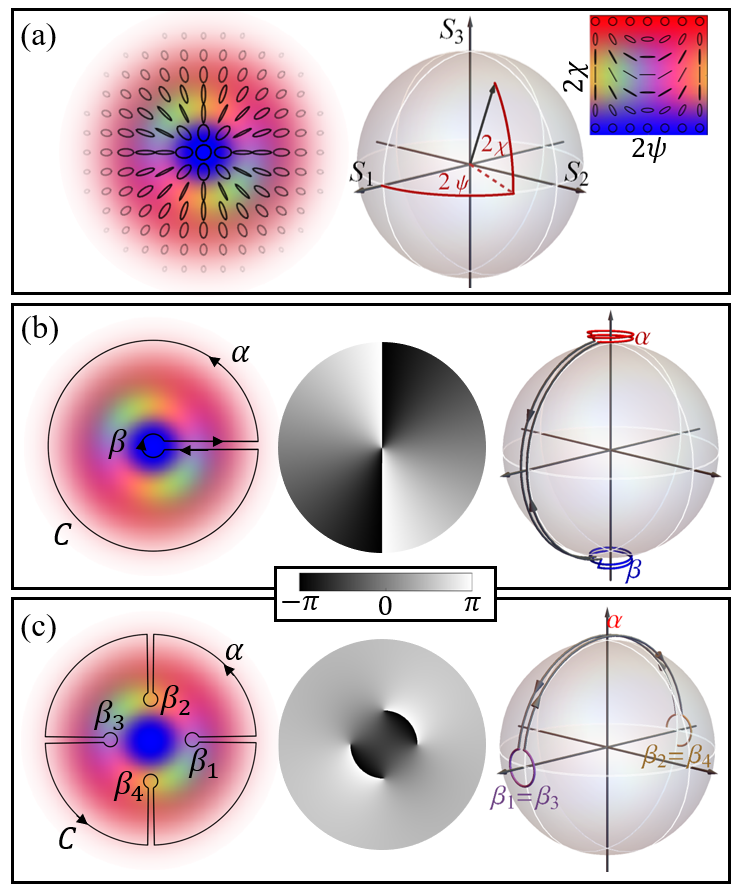}
    \caption{Topological measurement of the Skyrmion number. (a) Polarization profile of an $n=2$ Skyrmion and associated Poincar\'e sphere. Path integrals (left), phase profiles $\Phi$ (center) and Poincar\'e spheres (right) evaluated for $S_z=S_3$ in (b) and $S_z=S_1$ in (c).
    }
    \label{fig:line}
\end{figure}

The Skyrmion number can then be expressed as a line integral 
\begin{eqnarray}
    n =  -\frac{1}{4\pi}\oint_C S_z\nabla \Phi \cdot d\bm{l}, 
    \label{Eq.vIntegral2}
\end{eqnarray}
which depends only on the variation of the phase $\Phi$ along the path. 
The integration path has to enclose the entire (in principle infinitely extended) beam area, but exclude all singularities, as illustrated in \textbf{Figure~\ref{fig:line}}. The integrals connecting the beam periphery to the singularities %(labelled $\alpha$ and $\gamma$),
cancel. This leaves us with two (or more) closed line integrals: one at a radius $\rho \to \infty$, evaluated counter-clockwise ($\alpha$), and one or more around the singularities of $\Phi$, evaluated in clockwise direction ($\beta_j$): 
\begin{equation}
    n= \sum_j\frac{1}{4\pi}\oint_{\beta_j} S_z\nabla \Phi \cdot d\bm{l} -\frac{1}{4\pi}\oint_\alpha S_z\nabla \Phi \cdot d\bm{l}.\label{Eq.vIntegral3}
\end{equation}
At the positions of the $j^{\rm th}$ inner singularity $(x_j,y_j)$ the Stokes parameter is simply the local value $S_{z}^{(j)}$, and can be taken out of the integral.
At the beam periphery the Stokes parameter converges to a single value $\bar{S}_z^{(\infty)}$, because the Skyrmion beam carries a finite energy. 
Each integral is related to the winding number $N=(2\pi)^{-1}\oint \nabla \Phi \cdot d\bm{l}$, which counts the number of turns  $\bf{S}$ completes on the Poincar\'e sphere along the respective circular paths and by definition is free from noise. 
We thus obtain a topological definition of the Skyrmion number,
\begin{equation}
    n = \frac{1}{2}\left( \sum_{j} S_{z}^{(j)} N_j -\bar{S}_{z}^{(\infty)} N_\infty\right),
    \label{Eq.NewSkyrmion}
\end{equation}
which is a function of the Stokes vector at the position of the singularities and at the beam periphery, and the corresponding integer winding numbers on the Poincar\'e sphere. As $\bm{\Sigma}$
is invariant under rotations, the Skyrmion number $n$ does not depend on the orientation of the Poincar\'e sphere, giving us unlimited options of choosing $\vec{S}_R$ when calculating $n$.

We will now illustrate how Equation~(\ref{Eq.vIntegral2}) links to the topology of the polarization texture using the example of an $n=2$ Skyrmion  (of the form $\ket{\Psi_2}$ of Equation~(\ref{Eq:n}) introduced later in this letter).
Figure~\ref{fig:line}(a) shows the polarization texture, with local polarization states color-mapped to the Poincar\'e sphere as indicated in the inset. 
If we choose $\vec{S}_R=(S_1,S_2,S_3)$, as illustrated in Figure~\ref{fig:line}(b), the polarization profile features a left-handed C-point singularity in the center \cite{Dennis2009}, and a delocalized right-handed singularity at the beam periphery, as identified by the phase profile $\Phi$. A suitable path integral, indicated in the left, yields contributions from the line integrals around the central singularity ($S_3=-1$), and at $\rho \to \infty$ ($S_3=+1$).  
On the Poincar\'e sphere, these integrals correspond to winding twice backwards around the South and North pole, respectively, so Equation~(\ref{Eq.NewSkyrmion}) evaluates to $n=\frac{1}{2}((-1)\cdot (-2)-1\cdot (-2))=2$.  
Alternatively, we may choose $\vec{S}_R=(S_2,S_3,S_1)$ as illustrated in Figure~\ref{fig:line}(c). This results in four singularities: two horizontally and two vertically polarized.
The corresponding path on the Poincar\'e sphere is traversed twice, alternatingly winding around $S_1=1$ ($\beta_1$ and $\beta_3$) and $S_1=-1$ ($\beta_2$ and $\beta_4$), with positive and negative winding numbers respectively, and no contribution from the periphery, again resulting in the correct $n=2$.   

As anticipated, $n$ counts how many times the polarization wraps around the Poincar\'e sphere, taking into account the sense of the winding direction.
Unlike 
Equation~(\ref{Eq:SkyrmNum}) our topological expression requires neither derivatives nor integration but can be read directly from the polarization profile.
We will demonstrate in the following by experiment and simulation that our definition can provide a significant increase in accuracy and precision.

%%%%%%%%%%%%%%%%%%%%%%%%%%
\section{Experimental and numerical evaluation}
%%%%%%%%%%%%%%%%%%%%%%%%%%

When evaluating experimental data, the selection of the path is guided by fundamental as well as practical criteria: to obey Stokes's theorem, the enclosed area should contain as much as possible of the beam profile. 
For practical reasons, we want to choose a path which avoids areas of low intensity %at very large beam radii
as well as the immediate neighborhood of the singularities, where measurements of the Skyrmion field are dominated by noise. 

%-------------------------------------------------

\begin{figure}[htbp]
    \centering
    \includegraphics[width=16cm]{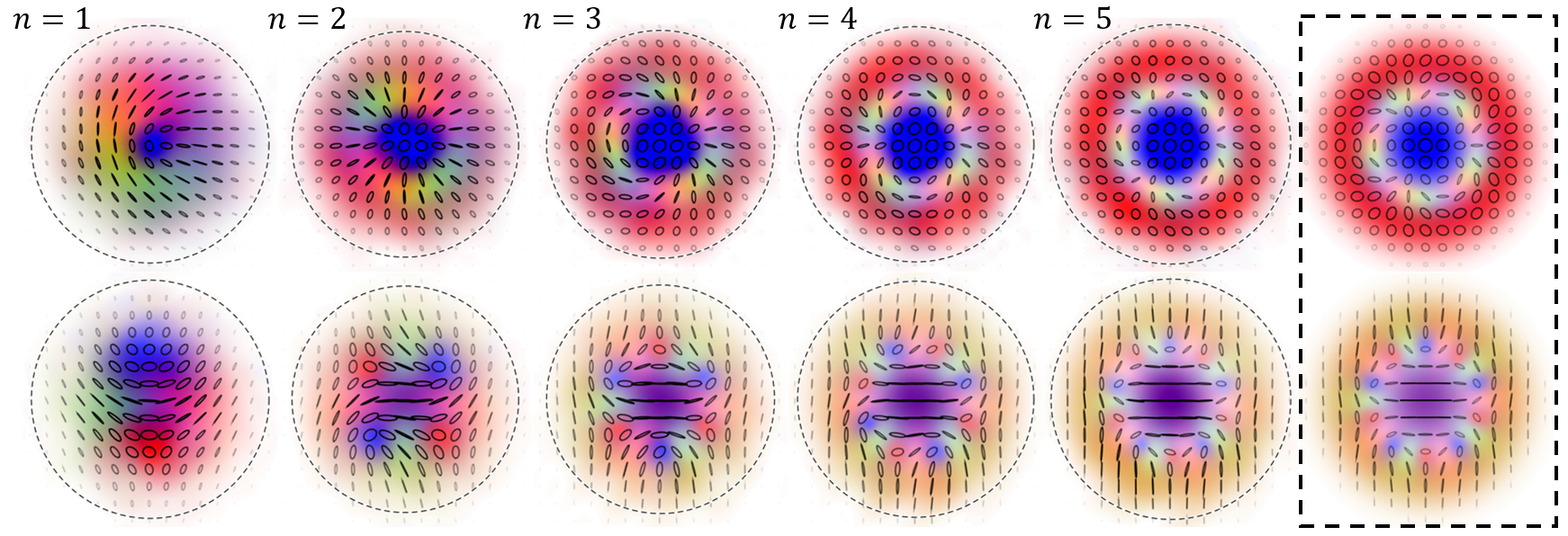}
    \caption{Experimentally measured polarization textures for beams with Skyrmion numbers $n=1$ to 5. The top row shows N\'eel-type Skyrmions defined in the circular basis ($\ket{0}=\ket{l}, \ket{1}=\ket{r}$), and the bottom row N\'eel-type bimerons in the linear basis ($\ket{0}=\ket{h}$ and $\ket{1}=\ket{v}$). The inset at the right shows the theoretical polarization textures of the corresponding target Skyrmions for $n=5$.}
    \label{fig:SkyrmionBeams}
\end{figure}

In the following we will compare measurement methods Equation~(\ref{Eq:SkyrmNum}) and Equation~(\ref{Eq.NewSkyrmion}) for Skyrmions and bimerons with various Skyrmion numbers, as well as for Skyrmion lattices.  
A Skyrmion with number $n$ is conveniently generated as a superposition of orthogonally polarized Laguerre-Gaussian (LG) modes $\textrm{LG}_p^\ell$ (full expression given in section A of the Supporting Information) \cite{Gao2020}: 
\begin{equation}
     \ket{\Psi_{n}} = \frac{1}{\sqrt{2}}\left( \textrm{LG}^0_0\ket{0}+\textrm{LG}^n_0\ket{1}\right).
     \label{Eq:n}
\end{equation}
Here $p$ and $\ell$ denote the radial and azimuthal mode order respectively, and $\ket{i}$ with $i \in \{0,1\}$ are any two orthogonal polarization states (i.e.~a Schmidt basis).  
Choosing circular (linear) polarization basis states results in N\'eel-type Skyrmions (bimerons). 

We generate vector beams of the form Equation~(\ref{Eq:n}) by encoding the LG modes 
as multiplexed holograms on a digital micromirror device (DMD), following the procedure outlined in \cite{Selyem2019,Rosales-Guzman2020} and detailed in Supporting Information B. To characterize our vector beam, we take spatially resolved Stokes measurements, which can be processed to obtain spatially resolved polarization profiles, as illustrated in \textbf{Figure~\ref{fig:SkyrmionBeams}} for Skyrmions and bimerons with $1 \leq n \leq 5$. These profiles show good qualitative agreement with the corresponding theoretical polarization profiles, as the reader can confirm by comparing the measured to the simulated $n=5$ target Skyrmions shown as inset in Figure~\ref{fig:SkyrmionBeams}. 

The spatially resolved Stokes measurements $\vec{S}(x,y)$ also form the basis of a quantitative analysis of $n$ according to the surface integral method Equation~(\ref{Eq:SkyrmNum}) and our topological method Equation~(\ref{Eq.NewSkyrmion}), summarised in Table \ref{table_1}. 
To ensure a fair comparison between the methods, the experimentally measured Stokes images were cropped to a disk across which the intensity is $\leq 5 \%$ of the peak intensity, as indicated by the dashed circles in Figure~\ref{fig:SkyrmionBeams}. Diffraction artefacts and noise were reduced by low-pass Fourier filtering the camera images. 

Equation~(\ref{Eq:SkyrmNum}) evaluates the Skyrmion number directly from the measured Stokes vectors $\vec{S}(x,y)$ and their numerical gradients.
The surface integral was performed over the entire grid space.

\begin{table}[b!]
    \centering
    \caption{Comparison of measured Skyrmion numbers for Figure~\ref{fig:SkyrmionBeams}
    evaluated using Equation~(\ref{Eq:SkyrmNum}) and Equation~(\ref{Eq.NewSkyrmion}) respectively.}
    \begin{tabular}{ccccccc}
    \hline
        type & method & $n=1$ & $n=2$ & $n=3$ & $n=4$ & $n=5$ \\
        \hline
        \multirow{3}{*}{Skyrmion}
         & Eq.~(\ref{Eq:SkyrmNum}) & 0.918 & 1.921 & 2.994 & 4.007 & 4.924 \\
          & Eq.~(\ref{Eq.NewSkyrmion}) ($S_z=S_3$) & 0.913 & 1.910 & 2.925 & 3.891 & 4.884 \\
          & Eq.~(\ref{Eq.NewSkyrmion}) ($S_z=S_1$) & 1.000 & 1.998 & 2.994 & 3.989 & 4.976 \\
        \hline 
        \multirow{3}{*}{Bimeron} & Eq.~(\ref{Eq:SkyrmNum}) & 0.927 & 1.941 & 2.971 & 3.999 & 4.991 \\
         & Eq.~(\ref{Eq.NewSkyrmion}) ($S_z=S_1$) & 0.915 & 1.931 &	2.970 & 3.966 & 4.972 \\
         & Eq.~(\ref{Eq.NewSkyrmion}) ($S_z=S_3$) & 1.000 & 1.998 & 2.992 & 3.989 & 4.993 \\
         \hline
    \end{tabular}
    \label{table_1}
\end{table}

Our topological method, Equation~(\ref{Eq.NewSkyrmion}), requires us to identify $\bar{S}_z^{(\infty)}$ (here taken as the edge of the disk), $S_z^{(j)}$ and the corresponding winding numbers.In Table~\ref{table_1} we present results for $S_z=S_3$ and $S_z=S_1,$ corresponding to the illustrations in Figure~\ref{fig:line}(b) and (c) respectively.

With both methods we obtain a Skyrmion number that closely matches the target value for each of the beams. We achieve the highest accuracy when evaluating Equation~(\ref{Eq.NewSkyrmion}) in a mutually unbiased basis to the Schmidt basis that defines the Skyrmion in Equation~(\ref{Eq:n}), i.e.~using $S_z=S_1$ for N\'eel type Skyrmions and $S_z=S_3$ for bimerons. This choice of generalized Stokes vectors shifts the relevant path integrals away from low intensity regions, where noise would compromise the evaluation of the Stokes parameters.

As the Skyrmion number is a global beam property, defined either by integration over an infinite transverse plane, or from evaluating the Stokes parameter at an infinite radius, any measurement is necessarily an approximation \cite{footnote2}. 
%Strictly speaking, this applies only to Skyrmions constructed from spatial modes that are defined over an infinite plane as e.g.~the LG modes used here or also Bessel modes.
Differences from the target Skyrmion number arise from inaccuracies in the experimental generation process as well as the numerical evaluation. An artefact of our particular Skyrmion `recipe' as defined in Equation~(\ref{Eq.NewSkyrmion}) is that, especially for Equation~(\ref{Eq:SkyrmNum}), accuracy improves for higher Skyrmion numbers, as the intensity profile of the two constituting spatial modes $\textrm{LG}^n_0$ and $\textrm{LG}^0_0$ overlaps less, so that $S_z^{(\infty)}$ is better defined.

Finally, we note that the obtained Skyrmion number is influenced by the extent to which filtering is performed. 
Additional detail on the experimental generation and analysis methods are presented in section B of the Supporting Information.

In our experiment we have generated Skyrmion beams with high fidelity, however in many situations one may not have this luxury, e.g.~when working at extremely low light levels, or when investigating light after propagation through noisy environments. It is perhaps no surprise that a topological identification of $n$ proves more 
effective to tackle noise, whereas noise amplification is an inherent property for the differentiation required in Equation~(\ref{Eq:SkyrmNum}).

\begin{figure}[htbp]
    \centering
    \includegraphics[width=8cm]{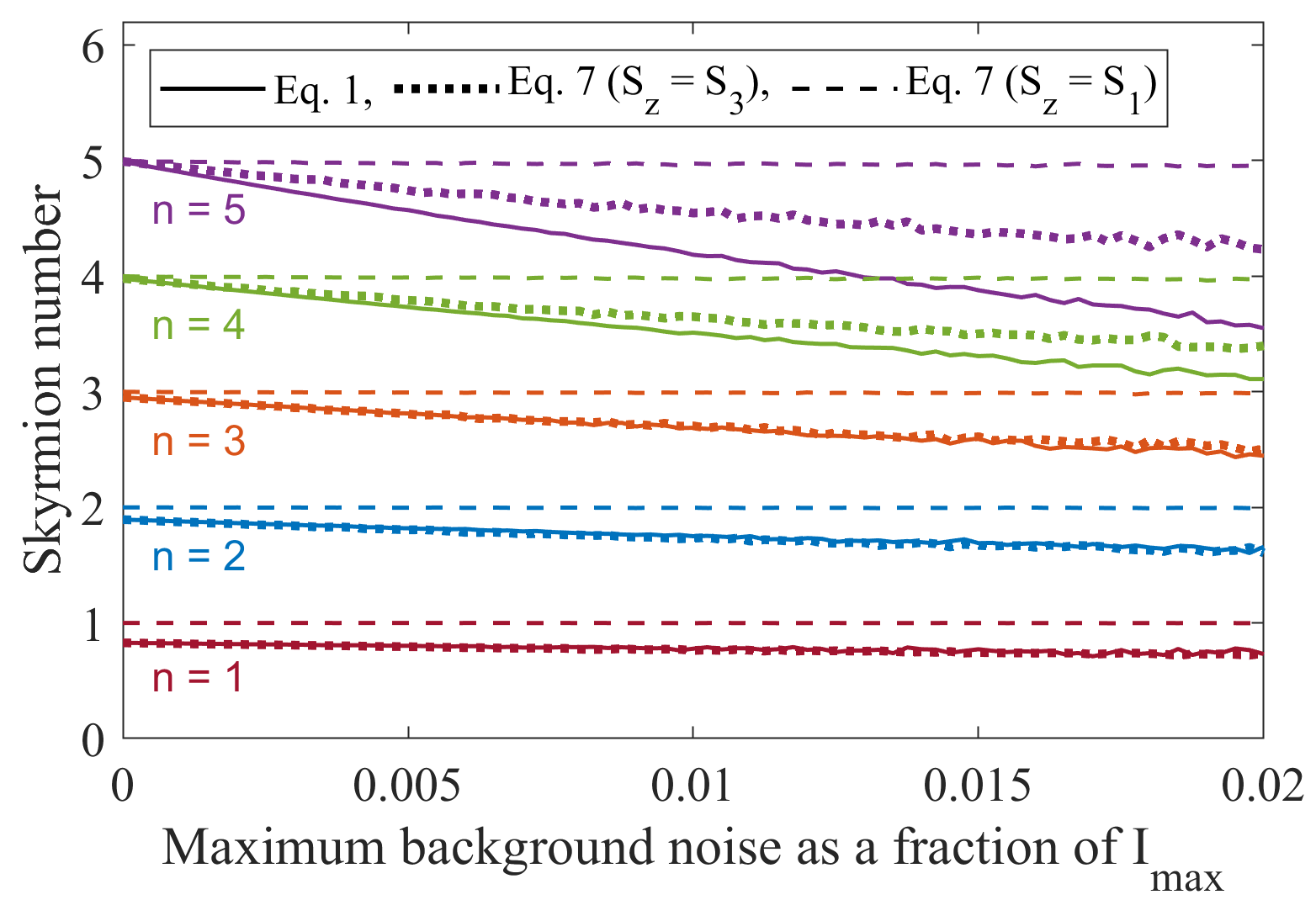}
    \caption{Comparison of Skyrmion measurement methods for increasing noise levels. Background noise is applied to simulated N\'eel-type Skyrmion beams with $n=1$ to 5, and evaluation is performed over the disk where intensity exceeds 5\% of the peak intensity. Solid lines indicate values calculated using Equation~(\ref{Eq:SkyrmNum}). Dotted lines are using Equation~(\ref{Eq.NewSkyrmion}) in the Schmidt basis of the beam ($S_z = S_3$) and dashed lines show values when evaluating in an orthogonal polarization basis ($S_z=S_1$).}
    \label{fig:NoiseComparison}
\end{figure}

We have confirmed this by applying artificial background noise to simulated data, adding random noise levels ranging from 0 to 2\% of the peak intensity $I_{\rm{max}}$ independently to each raw input image corresponding to the Stokes measurements. The resulting calculated N\'eel-type Skyrmion numbers for beams with $n=1$ to 5 (corresponding to the upper row of Figure~\ref{fig:SkyrmionBeams}) are shown in \textbf{Figure~\ref{fig:NoiseComparison}}.
We find that noise leads to an underestimation of $n$ for the integral method Equation~(\ref{Eq:SkyrmNum}) and less so for the topological method when evaluated in the Schmidt basis. Our simulations indicate that we can obtain an almost noise-free Skyrmion number (shown by dashed lines in Figure~\ref{fig:NoiseComparison}) when choosing to evaluate Equation~(\ref{Eq.NewSkyrmion}) using the mutually unbiased polarization basis. In addition, our simulations indicate an offset for low valued $n$ numbers, which again disappears for topological evaluation in the orthogonal polarization basis.

%%%%%%%%%%%%%%%%%%%%%%%%%%
\section{Characterization of Skyrmion Lattices}
%%%%%%%%%%%%%%%%%%%%%%%%%%

In the remainder of this letter we will highlight the geometric interpretation of Equation~(\ref{Eq.NewSkyrmion}) by discussing its application to Skyrmion lattices.  

\begin{figure}[htbp]
    \centering
    \includegraphics[width=9cm]{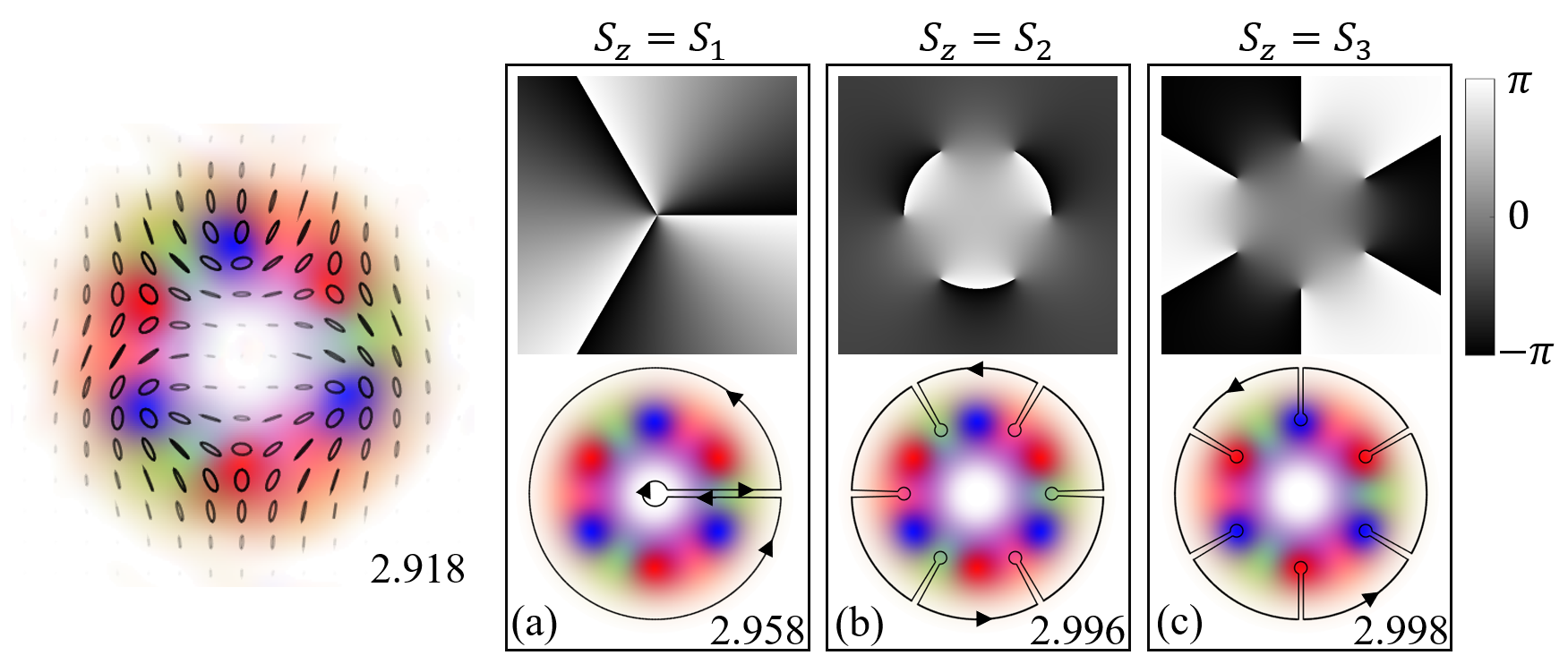}
    \caption{Interpretation of a Skyrmion lattice in terms of different polarization bases. Left: measured polarization texture of the Skyrmion lattice Equation~(\ref{Eq.Lattice}), displaying $n$ based on the integral method. Right: Phase $\Phi$ (top row) and integration path
    for $S_z=S_1,\ S_2,\ S_3$, respectively, displaying
   the corresponding Skyrmion number evaluated by the topological method. 
    }
    \label{fig:LatticeBeam}
\end{figure}

We have seen that for individual Skyrmions, a rotation of the Poincar\'e sphere leads to a different interpretation of the spin texture and its polarization singularities.  For Skyrmion lattices, it may also result in a different attribution of Skyrmion structures to the individual lattice sites.
We illustrate this with a Skyrmion ring lattice,  
\begin{equation}
    \ket{\Psi_L}=\frac{1}{\sqrt{2}} (\textrm{LG}^2_0 \ket{h}-\textrm{LG}^5_0\ket{v}). \label{Eq.Lattice}
\end{equation}
While such a ring lattice will be unfeasible in magnetic spin structures, optically we can generate more exotic Skyrmion textures. The measured polarization texture of this lattice is shown in the left of Figure~\ref{fig:LatticeBeam}. Based on the integral method (Equation~(\ref{Eq:SkyrmNum})) we obtain a Skyrmion number of 2.918, close to the overall target value of $n=3$.
The topological method yields even more accurate Skyrmion numbers of $n=2.958$, $n=2.996$ and $n=2.998$ for $S_z=S_1$, $S_2$ and $S_z=S_3$, respectively. The associated phase profiles and integration paths are provided in Figure~\ref{fig:LatticeBeam}(a-c).

For Figure~\ref{fig:LatticeBeam}(a), $S_z=S_1$ coincides with the Schmidt basis of the lattice as defined in Equation~(\ref{Eq.Lattice}) and the only singularity appears at the centre of the beam profile.
Figure~\ref{fig:LatticeBeam}(b) and (c) interprets the beam in terms of winding numbers around diagonal/antidiagonal and circular polarization singularities. The six singularities in the Stokes phase $\Phi$, where $S_z\to \pm 1$ each represent a meron which contributes a Skyrmion number of $1/2$, with no contribution from the beam edge. 
The experimentally obtained Skyrmion values obtained with $S_z=S_2$ and $S_3$ deviate from the ideal value of 3 only by about 1\textperthousand, an error reduced by an order of magnitude from that in the linear polarization basis. This arises from the higher accuracy in determining $S_z$ at singularities positioned in beam areas of higher intensity, which could be confirmed by applying artificial noise to a simulated ring lattice (details are provided in section C of the Supporting Information), providing further evidence that a judicious choice of the generalised Stokes basis allows us to optimise measurement protocols. A detailed investigation over evaluation areas with varying radii reported in section D of the Supporting Information shows that the topological method, if applied in an orthogonal basis, yields the correct $n$ as soon as all singularities are included within the evaluation area.)

\section{Conclusion}

In this letter, we have introduced a novel method for calculating the Skyrmion number of paraxial optical Skyrmion fields which captures the topological texture by line integration. While our method is expressed in terms of optical Skyrmions, the principles apply to any 2D Skyrmions, including e.g. magnetic Skyrmions in thin films.
We demonstrate excellent agreement between measured and theoretical Skyrmion numbers for various examples of experimental Skyrmion and bimeron beams constructed from superpositions of orthogonally polarized Laguerre-Gaussian modes. 
We demonstrate that our topological method, when evaluated for a suitably chosen basis, allows us to identify the correct Skyrmion number at noise levels where the integral method would fail.   
Our method also facilitates the exploration of novel topological structures beyond individual Skyrmions, as illustrated by the example of a Skyrmion lattice. 
Research on optical Skyrmions is still in its infancy. We anticipate that our new method to identify and assess the quality of experimental Skyrmion fields directly benefits the growth of the discipline.

% Experimental section

%\section{Experimental Section}
%\threesubsection{First part of experimental section}\\
%\threesubsection{Second part of experimental section}\\

% Acknowledgements
\medskip
\textbf{Acknowledgements} \par 
We thank Sijia Gao for early discussions on the nature and properties of optical Skyrmions.
A.M. acknowledges financial support from the UK Research and Innovation Council via grant EPSRC/DTP

2020/21/EP/T517896/1. C.M.C., S.M. B and S.F.-A. acknowledge financial support from the Royal Society through a Newton International fellowship NIF/R1/192384 and a Research Professorship RP150122. J.B.G acknowledges financial support from the EPSRC via grant EP/V048449/1 and the Leverhulme Trust.

% References
\medskip

% Use the following code if you wish to generate your bibliography with BibTeX;
% replace the string "MSP-template" below with the name(s) of
% the BibTeX data base(s) you want to use.
% The resulting bibliography-output (the content of the .bbl file)
% must be pasted back into this file before submission.
% Please also include your BibTeX data base file(s) in your submission
% so that we can re-run BibTeX if necessary.
%

%\bibliographystyle{MSP}
%\bibliography{bibliography}

\begin{thebibliography}{10}
\bibitem{footnote1}
Note that while every Skyrmion beam is a Poincar\'e beam, the reverse does not
  hold: The mapping must obey specific mapping rules. This is intrinsic to the
  definition of the Skyrmion number, Equation.~\ref{Eq:SkyrmNum} but becomes
  more evident from our geometric description in this letter.
\bibitem{footnote2}
Strictly speaking this applies only to Skrymions constructed from spatial modes
  that are defined over an infinite plane as e.g.~the LG modes used here or
  also Bessel modes.
\end{thebibliography}

\providecommand{\noopsort}[1]{}\providecommand{\singleletter}[1]{#1}%

% Figures/tables and captions
% Permission statements are required for all figures reproduced or adapted from previously published articles/sources. Please also ensure that all necessary permissions to reproduce images have been received
% Please remove these statements for original figures

\section*{Supplementary material}

In this supplementary material, we provide expressions for Laguerre Gaussian beam modes, full experimental details for the generation of Skyrmion beams, including descriptions of the processing and analysis of measurements, and discussions relating to the application of artificial noise to simulated data. 

\subsection*{I. LG beam modes}

A Laguerre-Gaussian (LG) mode of order $l, p$, expressed in cylindrical polar coordinates, $r = \sqrt{x^2+y^2}$, $\phi = \arctan(y/x)$, and for a propagation distance $z$, is given by,
\begin{equation}
\begin{split}
   \textrm{LG}^\ell_p(r,\phi,z) =& \sqrt{\frac{2 p!}{\pi  (| \ell| +p)!}}\frac{1}{w(z)} \left[\frac{r \sqrt{2}}{w(z)} \right]^{|\ell|}L^{|\ell|}_p\left(\frac{2 r^2}{w(z)^2}\right) \\
   & \times \exp\left(-\frac{r^2}{w^2(z)}\right)  \exp(i(\ell \phi+ \Phi))\\
   & \times \exp \left(i\frac{k r^2 z}{2(z_\mathrm{R}^2+z^2)}\right),
   \end{split}
\end{equation}
where, $z_\textrm{R}=\pi w^2_0/\lambda$ is the Rayleigh range, $w=w_0 (1+\left(z/z_\mathrm{R}\right)^2)^{1/2}$ is the beam radius and $w_0$ is the waist radius. 
The Gouy phase for a mode of order $N=|\ell|+2p$, is $\Phi = (N+1)\arctan\left(z/z_R \right)$. $L^{\vert \ell\vert}_p$ are Laguerre polynomials.

\subsection*{II. Experiment details}

\begin{figure}[b]
    \centering
    \includegraphics[width=0.95\linewidth]{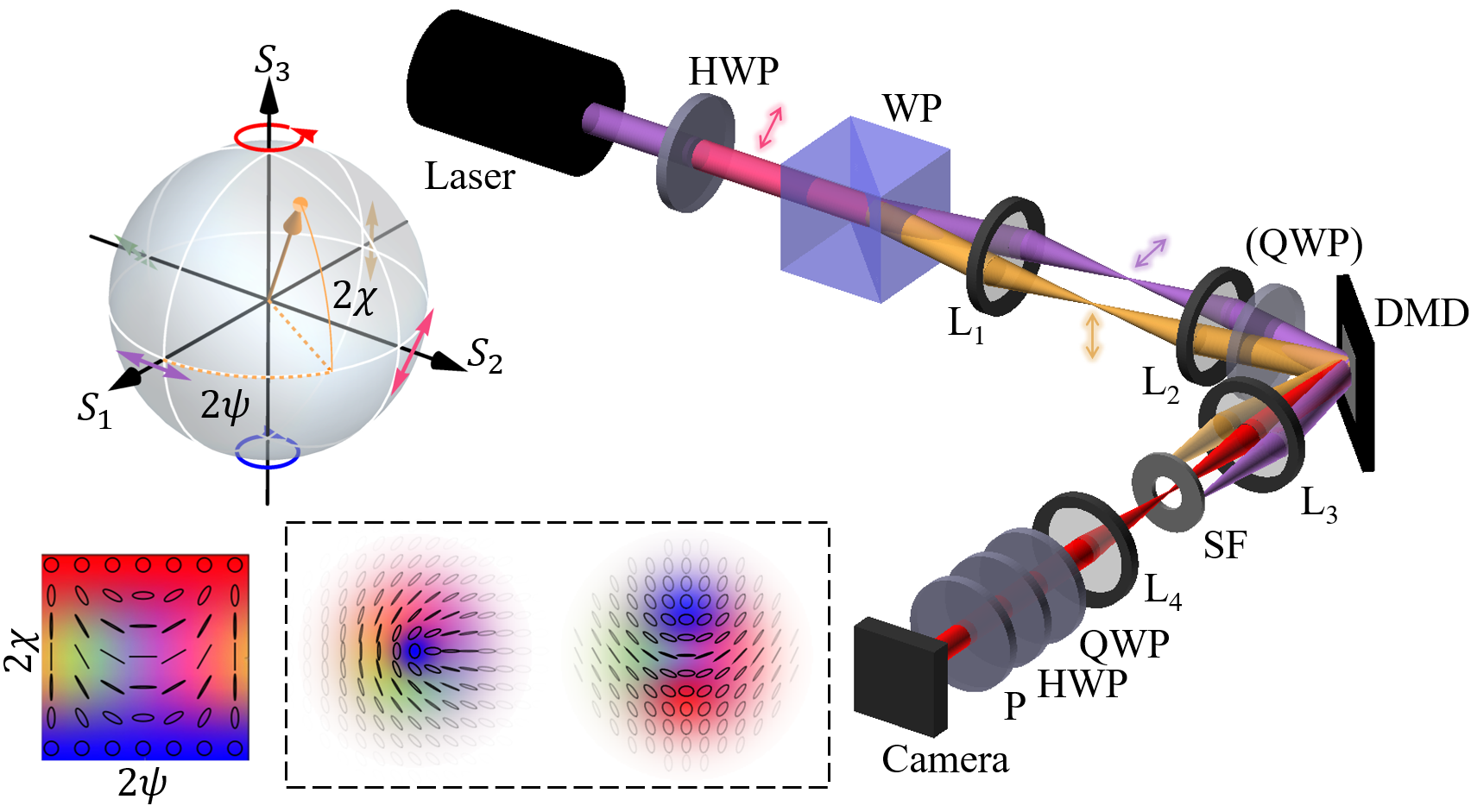}
    \caption{Experimental generation of optical Skyrmions and bimerons. Left: Poincar\'e sphere, defining the Stokes vector $\vec{S}$ and indicating the color map for polarization states used throughout this paper. Right: experimental setup. Inset: experimental polarization profiles of a $n=1$ Skyrmion beam (left) and a $n=1$ bimeron beam (right) generated using the
    right-left handed circular polarization basis and the horizontal-vertical polarization basis, respectively.}
    \label{fig:SetUp}
\end{figure}

We employ a digital micromirror device (DMD) to generate  vector beams including bimerons and Skyrmions.  
A schematic of the experimental setup is shown in Figure~\ref{fig:SetUp}. 
Here, a collimated, horizontally polarized laser beam ($\lambda=633$nm) propagates through a half wave-plate (HWP), creating diagonally polarized light. A Wollaston prism (WP) separates the beam into its horizontal and vertical components, with an angular separation of $1^\circ\: 20^\prime$. A telescope formed of lenses $\rm{L}_1$ and $\rm{L}_2$ overlaps the beams 
onto the DMD, on which a multiplexed hologram is displayed. As the beams are incident on the DMD at different angles, the hologram can shape the amplitude and phase of each beam independently and ensures that the first diffraction orders of the orthogonal beams overlap, where the vector beam is generated. 
A telescope ($\rm{L}_3$, $\rm{L}_4$) and a spatial filter (SF) are used to block unwanted diffraction orders. A quarter wave-plate (QWP) can be added before the DMD to obtain vector beams using the circular polarization basis. The intensity of the generated beam is recorded using a CMOS camera. Spatially resolved Stokes measurements are preformed using a fixed linear polarizer (P) and a rotating QWP and HWP.
The Stokes parameters are obtained from intensity images corresponding to horizontal, vertical, diagonal, antidiagonal, right and left circular polarizations ($\hat{h}, \hat{v}, \hat{d}, \hat{a}, \hat{r}$ and $\hat{l}$), according to:
\begin{equation}
\begin{split}
    S_0 &= I=I_h+I_v = I_d+I_a = I_r+I_l, \\
    \vec{S}&=(S_0,S_1,S_2,S_3)^T \\
    &= (I,I_h-I_v, I_d-I_a, I_r-I_l)^T.
\end{split}
\end{equation}
Background light intensity, of approximately 1.5\% of the peak intensity, is subtracted from each of the six intensity measurements. Its value is determined by averaging over a small area far from the beam. 
We apply a low pass Fourier filter to each intensity image, using a kernel with a top hat profile, to remove artefacts due to diffraction.
Theoretically, the expressions for $S_0 = I$ are equivalent, but the measured values may differ slightly depending on the chosen basis.
Consequently, normalised Stokes parameters are obtained by dividing by the total intensity obtained from the sum of the corresponding polarization states.

When using Equation~1 to measure the Skyrmion number, we perform the surface integral over a disk, centred on the beam origin. For the values obtained in Table I, the radius of this disk was chosen such that the intensity at the edge falls to 5\% of the peak intensity. 

Measuring the Skyrmion number using Equation~7 requires the evaluation of $S_z$ at the location of each singularity. To do so, we perform an average over a square of $3\times3$ pixels, centred on the location of the singularity, with the pixel size given by the camera pixel size, $5.2 \ \mathrm{\mu m}$.
We calculate the value of $\bar{S}_z^{(\infty)}$ by averaging the values of $S_z$ lying along a circular path, centred on the beam centre. For the Skyrmion numbers given in Table I, this circular path was again chosen to lie at where the beam intensity falls to 5\% of the peak intensity.

The winding numbers $N$ in Equation~7. describe the number of turns completed by $\vec{S}$ on the Poincer\'e sphere along a path surrounding a singularity. Plotting the Stokes phase $\Phi=\arctan(S_y/S_x)$ of the complex Stokes field, $ S_\pm = S_x \pm i S_y = |S_\pm|e^{\pm i\Phi}$, visually gives the location of each singularity and the corresponding winding numbers.
Computationally, the singularities can be found at the positions where $S_z \to \pm1$, or equivalently, where $S_x$ and $S_y \to 0$. 
Converting a small area about the $j^{th}$ singularity into a polar plot allows us to obtain the magnitude of the winding number $N_j$ by counting the peaks in the angular direction of the polar plots, with the sign of the gradient giving the sign of $N_j$. 
For experimental data, $N$ is calculated by averaging over 10 rows of the polar plot.
To find $N_\infty$, the entire grid space is converted into a polar plot and we count the peaks near the edge of the beam profile.  

\subsection*{III. Simulated response of the various methods to noise}

\begin{figure}[t]
    \centering
    %\vspace{2mm}
    \includegraphics[width=0.95\linewidth]{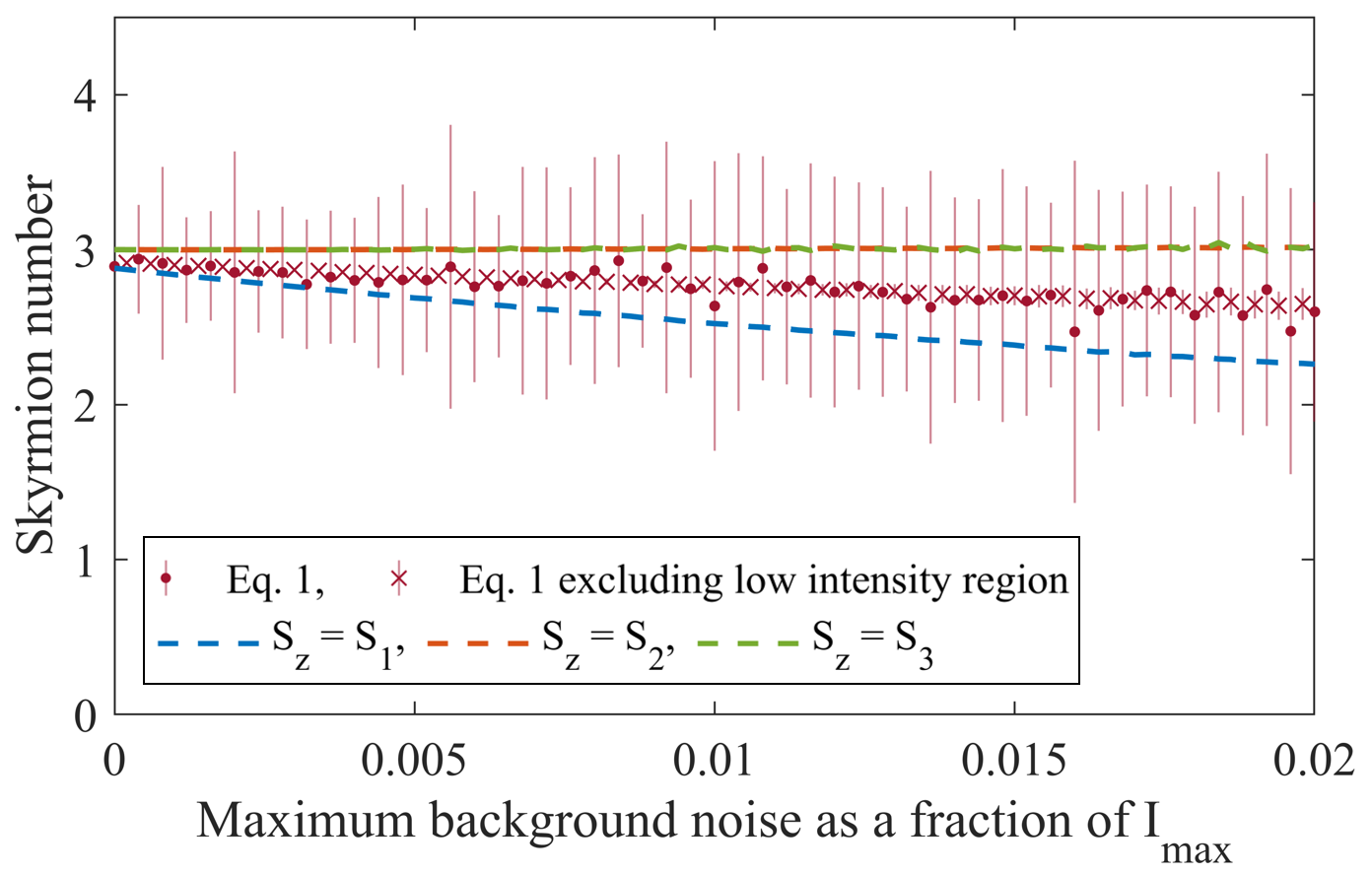}
    \caption{Comparison of Skyrmion measurement methods for the ring lattice beam shown in Figure~4 to simulated increasing background noise levels. Red points show the average of 100 $n$ values calculated using Equation~1 directly, whereas red crosses show $n$ calculated using Equation~1, but omitting the central region of the beam where intensity is below 0.05$I_{\text{max}}$. Error bars indicate the standard deviation. Dashed lines show $n$ evaluated using Equation~7, for the Schmidt basis of the beam (blue), and the two orthogonal polarisation bases (orange and green). The standard deviation when using Equation~7 was calculated to be less than 0.04 for the maximum noise level.}
    \label{fig:LatticeNoise}
\end{figure}

Regions of low intensity directly affect the Skyrmion number as the polarization at the beam periphery can determine whether the mapping of the Poincar\'e sphere is complete over the transverse profile. The measured Skyrmion number thus deteriorates in the presence of noise in these regions.
In the main paper, we showed how the various methods of calculating $n$ responded to artificially applied background noise to simulated N\'eel-type Skyrmion beams. Here we provide additional insights by showing the response of the different methods to a Skyrmion lattice with applied noise. 

For the same ring lattice beam described by Equation~9, we apply background noise by adding random values ranging between 0 and $\alpha I_{\rm{max}}$ to simulated ($\hat{h}, \hat{v}, \hat{d}, \hat{a}, \hat{r}$, $\hat{l}$) intensity measurements. Here, $I_{\rm{max}}$ is the maximum intensity in the entire beam, and $\alpha$ is a parameter controlling the noise level. 
Figure~\ref{fig:LatticeNoise} shows the calculated Skyrmion numbers for increasing levels of background noise ($0\le\alpha\le 0.02$), when evaluating using either Equation~1, or Equation~7  with $S_z=S_1,\ S_2,\ S_3$. 
We apply each noise level 100 times, calculate $n$ each time, and display the average of the obtained values in Figure~\ref{fig:LatticeNoise}, with the standard deviation shown as error bars.

For each method, $n$ is evaluated where the intensity falls to 5\% of the maximum beam intensity. 
From the circle data points, it can be seen that the surface integral method can greatly under- or overestimate the Skyrmion number, with a large standard deviation over the 100 iterations. This large variation can be put down to the low intensity region in the centre of our ring lattice. However, when the central low intensity region is excluded from the integration (shown by crosses) $n$ still deteriorates with increased noise. When the topological method is performed in the Schmidt basis of the beam (blue dashed line), the measured $n$ was found to also deteriorate with increased noise. The true advantage of of Equation~7 is seen when choosing to evaluate using an orthogonal polarisation basis to the Schmidt basis (orange and green dashed lines). Here, we have been able to correctly identify the Skyrmion number of the lattice for all shown noise levels. 

Our simulations indicate that the topological method is advantageous when dealing with noisy experimental data for a suitable choice of $S_z$.

\subsection*{IV. Variation of Skyrmion number with evaluation area}

\begin{figure}
    \centering
    \includegraphics[width=0.95\linewidth]{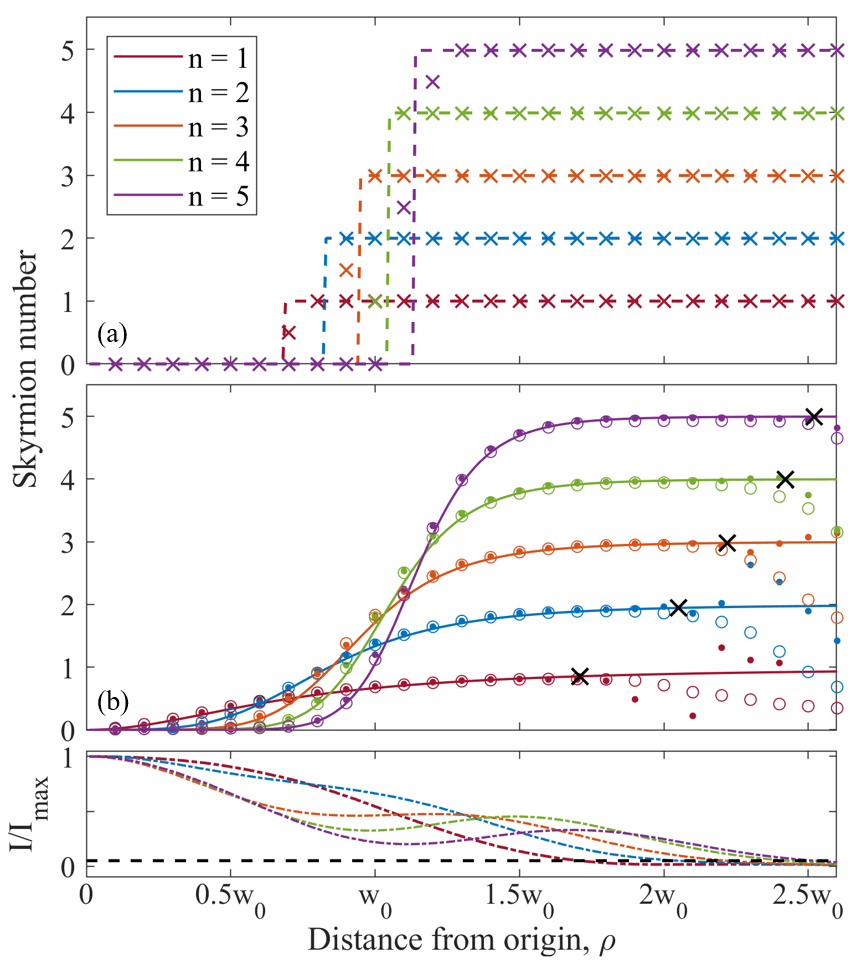}
    \caption{Comparison of Skyrmion numbers obtained from the integral and topological definition evaluated over varying radii $\rho$. We evaluate N\'eel Skyrmions of the form Equation~8
    for Skyrmion numbers $n=1$ to 5.
    The top plot shows simulated (dashed lines) and experimentally (crosses) obtained Skyrmion numbers when using the topological method and evaluating in an orthogonal polarisation basis ($S_z = S_1$).
    The middle plot compares the two methods when choosing $S_z$ to correspond to the Schmidt basis ($S_z = S_3$).  
    Simulations are shown as  solid  lines,  and experimental data as closed (Equation~1) and open (Equation~7) data points. 
    The lower plot shows the peak-normalised beam intensity of the experimentally measured beams. The dashed line marks 5\% of $I_{\rm{max}}$, and the beam radii where the various beams reach this intensity are indicated by black crosses in the middle plot.}
    \label{fig:MethodComparisonPlot}
\end{figure}

%As the Skyrmion number is a global beam property, defined by integration over an infinite transverse plane, or from evaluating the Stokes parameter at an infinite radius, any measurement is necessarily an approximation.
%\footnote{Strictly speaking this applies only to Skrymions constructed from spatial modes that are defined over an infinite plane as e.g. the LG modes used here or also Bessel modes}.
The Skyrmion number is a global property of a beam which relies on evaluation over an infinite transverse plane. As such, any measurement will be an approximation.  
We illustrate this by evaluating the numerical and measured Skyrmion number over a restricted circular region of the transverse plane, varying the radii $\rho$ from 0.1 up to 2.5 times the beam waist $w_0$. In Figure~\ref{fig:MethodComparisonPlot}, we compare the response of the surface integral and topological methods. 

In Figure~\ref{fig:MethodComparisonPlot}(a), the topological method, Equation~7, has been used to evaluate N\'eel type Skyrmion beams with $n=1$ to 5, with simulation shown as dashed lines and experimental measurements as crosses. Here, $S_z = S_1$ was chosen, resulting in the calculation being performed using a polarisation basis orthogonal to the Schmidt basis of the beam. From this, it is clear that the measured Skyrmion number remains zero for lower radii, until all the singularities are included within the evaluation area, when the correct Skyrmion number is identified. The experimental results match the simulation with very little deviation. 

In Figure~\ref{fig:MethodComparisonPlot}(b) we show the same variation of the Skyrmion number to increasing evaluation radii, however, this time using the surface integral method (Equation~1) and the choosing to evaluate the topological method using the Schmidt basis ($S_z=S_3$). 
Simulations are shown as solid lines, and the corresponding measured data using Equation~1 (Equation~7, in the Schmidt basis of the beam) are included as solid (open) data points. The simulations indicate that the obtained Skyrmion numbers indeed approach their target values for increasing beam areas and that simulations based on  Eqs.~1 and 7 are indistinguishable. 

For both techniques, the experimental results initially follow  simulation and approach the expected Skyrmion number. Beyond certain values of $\rho$, however, measurements are compromised by noise in the low intensity regions: the integration method  Equation~1 produces fluctuating Skyrmion numbers, and for the topological method Equation~7
they gradually decrease. Measurements begin to deviate from simulation for a beam radius $\rho$ where the intensity is decreased to roughly 5\% of the peak intensity, $I_{\rm{max}}$. For this reason, all the experimentally measured Skyrmion numbers stated in Table I were calculated using data cropped to a disk, outside of which the intensity falls below 5\% of $I_\text{max}$.

Our investigation of Figure~\ref{fig:MethodComparisonPlot}(b) shows that the topological method response to noise is more consistent, and the Skyrmion number is always underestimated in areas dominated by noise, due to $\big|\lim\limits_{\rho\to\infty}\bar{S_z}(\rho)\big| \leq 1$. The true value of $n$ may therefore be obtained by identifying the radius where $n$ is maximal, and then rounding to the nearest integer.

\end{document}